\documentclass[%
 reprint,prb,
 amsmath,amssymb,
 aps,
floatfix,superscriptaddress
]{revtex4-2}
\usepackage{comment}
\usepackage{hyperref}
\usepackage{graphicx}
\usepackage{dcolumn}
\usepackage{bm}
\usepackage{color}
\usepackage{xcolor}
\usepackage{soul}

\usepackage[caption=false]{subfig}
\usepackage{appendix}
\usepackage{chngcntr}

\hypersetup{
    colorlinks=true,   
    linkcolor=black,   
    citecolor=blue,    
    urlcolor=blue,     
}

\begin{document}

\preprint{APS/123-QED}

\title{Non-equilibrium correlation effects in spin transport through the 2D ferromagnet Fe$_4$GeTe$_2$}

\author{Declan Nell}
\affiliation{School of Physics and CRANN Institute, Trinity College Dublin, The University of Dublin, Dublin 2, Ireland}
\author{Stefano Sanvito}
\affiliation{School of Physics and CRANN Institute, Trinity College Dublin, The University of Dublin, Dublin 2, Ireland}
\author{Andrea Droghetti}
\email[]{andrea.droghetti@unive.it}
\affiliation{Department of Molecular Sciences and Nanosystems, Ca' Foscari University of Venice, via Torino 155, 30170, Mestre, Venice, Italy}

\begin{abstract}

Understanding non-equilibrium spin transport through 2D ferromagnets is a theoretical challenge, as correlations produce a complex electronic structure with coexisting itinerant and localized electrons.
We have developed a fully non-equilibrium {\it ab initio} method, combining density functional theory, 
dynamical mean-field theory, and non-equilibrium Green's functions to investigate the transport in 
Fe$_4$GeTe$_2$, a prototypical high-temperature 2D ferromagnet. 
We show that, while spin transport remains essentially single-particle under moderate bias, inelastic spin-dependent scattering of carriers with particle–hole excitations drives a distinctive hot-correlated electron regime beyond a critical voltage. This regime is marked by incoherent features in both the electronic spectrum and the conductance, which are experimentally accessible. Our results  demonstrates that material-specific many-body non-equilibrium methods are essential for a complete understanding of spin transport in 2D ferromagnets.
\end{abstract}

\maketitle

\textit{Introduction.} Electron correlations are central to spintronic materials, particularly in the emerging class of van der Waals (vdW) ferromagnets. The Fe$_n$Ge(Ga)Te$_2$ ($n=3$–$5$, FGT) family exemplifies this behavior \cite{Gong2017,Huang2017,2dmagnets1,2dmagnets2,first_FeGaT_paper}: its magnetism arises from the interplay between itinerant and localized Fe-$3d$ electrons \cite{nonstonerFGT,Ch.Ya.13,ve.ts.15,zh.lu.18}, leading to robust ferromagnetism that persists to the monolayer limit and above room temperature \cite{zh.ja.16,bsanyal}. These properties make FGT a promising platform for high-performance magnetic tunnel junctions \cite{tsymbol-mtj,Tsymbol_mtjhbn,FGT-mtj-acs,FGT3-pra-mtj,FGT-mtj-mol,FGT4-bsanyal,hBN,hbn2,MoS2,InSe,GaSe,WSe2,WS,FGaT-WSe2exp,Parkin-et-al}.

Despite the crucial role of correlations, theoretical studies of FGT-based devices typically rely on coherent electron transport through spin-polarized single-particle states \cite{bu.zh.01,ma.um.01}, most often within Kohn–Sham (KS) density functional theory (DFT) \cite{jo.gu.89,kohn.99,jone.15}. Going beyond this picture remains a major challenge. Existing studies of correlated transport have been confined to linear-response~\cite{andrea_Cu_co,liviu_Cu_Co_dmft,mo.ap.17,ha.ne.24,Ab.Mr.24,ne.dr.25}, while finite-bias conditions 
have been addressed only in simplified one-band models \cite{ok.07,ok.08,ar.kn.13}.

\begin{figure*}[t]
\centering\includegraphics[width=1\textwidth]{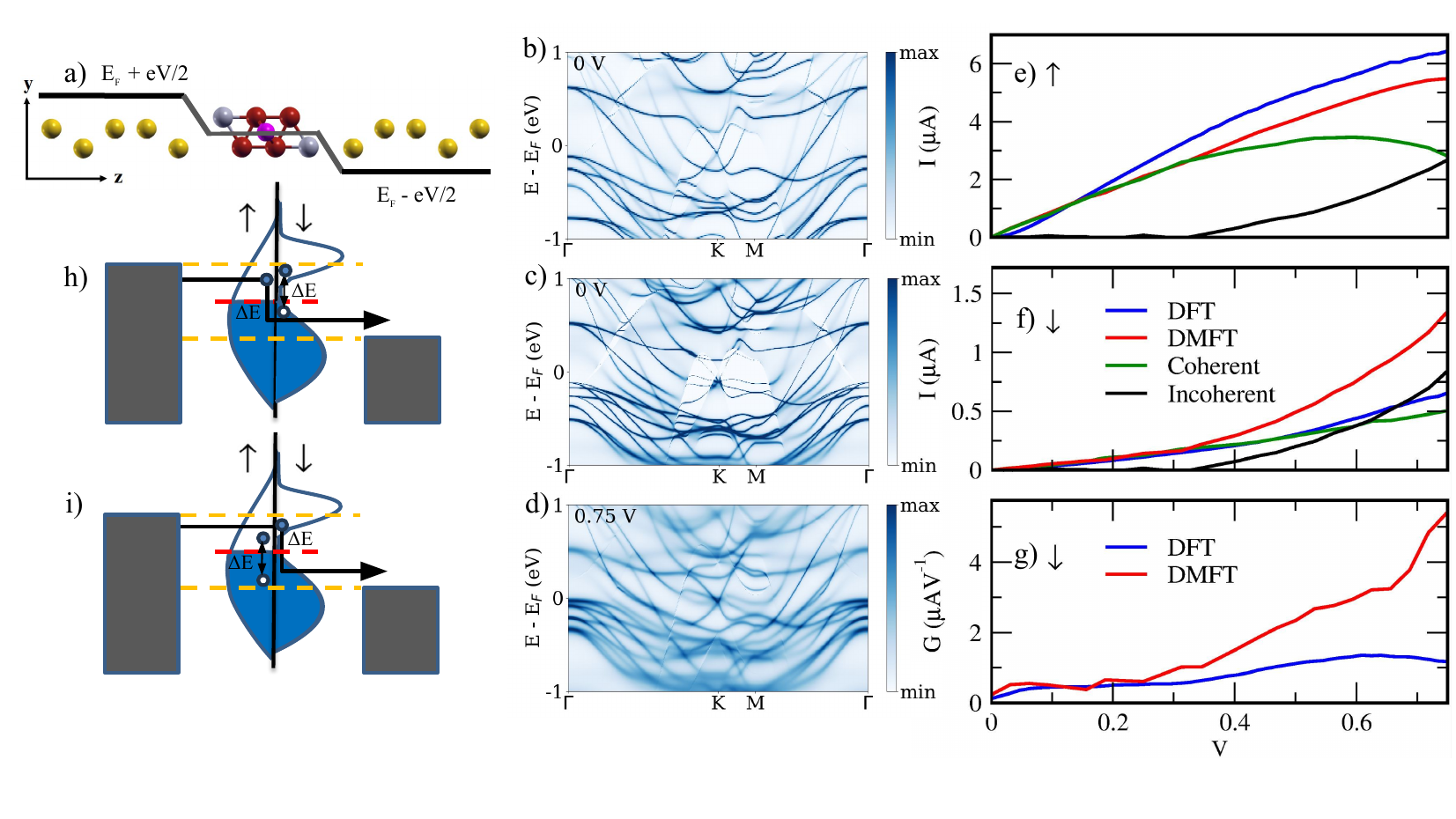}
{(a) Atomic structure of the Fe$_4$GeTe$_2$ monolayer device, where Fe, Te, Ge, and electrode atoms are shown as red, gray, magenta, and gold spheres, respectively. The schematic potential profile under an applied bias is superimposed across the device. (b,c) $\boldsymbol{k}$-resolved Fe $3d$ spectral functions from DFT and DMFT at equilibrium. The apparent discontinuity, visible as a light region in the center of the spectrum, arises from
hybridization with the bands of the electrodes. (d) DMFT spectral function at 0.75 V. (e,f) Spin-resolved currents versus bias. (g) Spin-down conductance versus bias. (h,i) Cartoon of dominant inelastic scattering for 
spin-up and spin-down electrons. Dark gray rectangles represent the Fermi boxes of the electrodes with the schematic Fe PDOS shown in between. Filled PDOS are shaded in blue. The red line indicates $E_\mathrm{F}$, while the orange 
lines mark the bias window.  Electrons and holes are represented 
by filled and empty dots, respectively.}\label{fig: spectral_function}
\end{figure*}

Here we introduce a combination of DFT, dynamical mean-field theory (DMFT) \cite{me.vo.89,ge.ko.96} and non-equilibrium 
Green's functions (NEGF)~\cite{bookStefanucci,Da.95,book1,Ta.Gu.01,Ba.Mo.02,ro.ga.06} to 
study the spin-dependent properties of Fe$_4$GeTe$_2$ monolayer [Fig.~\ref{fig: spectral_function}(a)] at finite bias, thereby going beyond the limitations of previous
studies. 
We find that at zero-bias Fe$_4$GeTe$_2$ has Fermi-liquid character and acts as an almost perfect 
spin filter, meaning that only electrons in the majority spin channel conduct, like in a half-metal. 
At higher bias, however, the voltage enhances inelastic scattering of carriers with collective electron-hole excitations [Fig. \ref{fig: spectral_function}(h) and (i)], driving the systems into a regime of ``hot-correlated electrons''.
This is characterised by incoherent features in both the electronic spectrum [Fig.~\ref{fig: spectral_function}(d)]  and the conductance [Fig.~\ref{fig: spectral_function}(d)], which are experimentally accessible and potentially relevant for devices performance. Beyond Fe$_4$GeTe$_2$, 
similar behavior may occur in other vdW ferromagnets, highlighting the crucial role of bias-driven non-equilibrium correlations in 2D spintronics.

\textit{Device setup.} Our prototypical device [see Fig.~\ref{fig: spectral_function}(a)] consists of an 
Fe$_4$GeTe$_2$ monolayer contacted to two semi-infinite metallic electrodes~\cite{ha.ne.24}
in a two-probe setup. 
Transport occurs along the $z$-axis, namely perpendicular to the monolayer, 
while periodic boundary conditions are applied in the transverse plane. The device is partitioned into left (L) and right (R) leads and a central region (CR), represented within a numerical atomic orbital basis set~\cite{siesta}. The leads are maintained at their local chemical potentials, $\mu_{\textnormal{L,R}}$,
which in equilibrium define a common Fermi energy, $E_\mathrm{F}\equiv\mu_\textnormal{L}=\mu_\textnormal{R}$.
In contrast, a bias voltage, $V$, shifts the local chemical potentials, $\mu_{\textnormal{L}(\textnormal{R})}=E_\mathrm{F}\pm eV/2$,
and drives the system out of equilibrium inducing current flow.

\textit{Theoretical framework.} 
The NEGF of the CR, $G^\sigma$, for spin $\sigma (= \uparrow, \downarrow$) is a $2\times2$ 
block matrix~\cite{ra.sm.86, al.06, ar.kn.13} containing the retarded, $G^{\sigma\;r}$, and advanced,
$G^{\sigma\;a}$, Green's functions on the diagonal and the Keldysh one, $G^{\sigma\;K}=G^{\sigma\;r}-G^{\sigma\;a} +2 G^{\sigma\;<}$, in the upper off-diagonal
position, where
$G^{\sigma\;<}$ is the lesser Green's function (see the End Matter - EM). At steady-state, 
the energy-dependent $G^\sigma$ satisfies Dyson's equation
\begin{equation}\label{eq: dyson}
G^{\sigma}(E,\boldsymbol{k})=g^{\sigma}(E,\boldsymbol{k})+g^{\sigma}(E,\boldsymbol{k})\Sigma^{\sigma}(E,\boldsymbol{k})G^{\sigma}(E,\boldsymbol{k}).
\end{equation}
where $g^\sigma(E,\boldsymbol{k})$ is the NEGF of the single-particle KS Hamiltonian of the CR 
and $\Sigma^{\sigma }(E, \boldsymbol{k})$ is the self-energy. Here, $\boldsymbol{k}$ is the transverse 
wave-vector over the 2D Brillouin zone (BZ). $A^\sigma(E,\boldsymbol{k}) =  i\big[G^{\sigma\; r}(E,\boldsymbol{k}) - G^{\sigma\; a}(E,\boldsymbol{k})\big] $ defines the spectral function, whose integral over the BZ returns the density of states (DOS). 

$\Sigma^{\sigma }(E, \boldsymbol{k})$ comprises two contributions: the leads' self-energies, $ \Sigma^{\sigma}_\mathrm{L,R}(E,\boldsymbol{k})$, 
describing the coupling of the CR with the leads, and the many-body self-energy 
$ \Sigma^{\sigma}_\mathrm{MB}(E,\boldsymbol{k})$, capturing the effect of electron-electron interaction. 
These have the same block form of $G^\sigma$ and can be decomposed into retarded, advanced 
and lesser components. The real and imaginary parts of $\Sigma^{\sigma \;  r}(E,\boldsymbol{k})$ account 
for the level renormalization and spectral broadening, respectively, while $\Sigma^{\sigma \; <}(E,\mathbf{k})$ 
describes the charge inflow and the interaction-induced changes in electron occupation.
For $\Sigma^{\sigma}_\mathrm{MB}(E,\boldsymbol{k})=0$, Eq.~(\ref{eq: dyson}) reduces to the standard 
DFT+NEGF formalism \cite{Ta.Gu.01,Ba.Mo.02,ro.ga.06}, providing a static mean-field description of the 
interaction. Here we go beyond such picture and compute $\Sigma^{\sigma}_\mathrm{MB}(E,\boldsymbol{k})$, extending the DMFT framework introduced in \cite{andrea_sigma_2,andrea_Cu_co} to the finite-bias 
limit. 

\begin{figure*}[t]
\centering\includegraphics[width=0.8\textwidth]{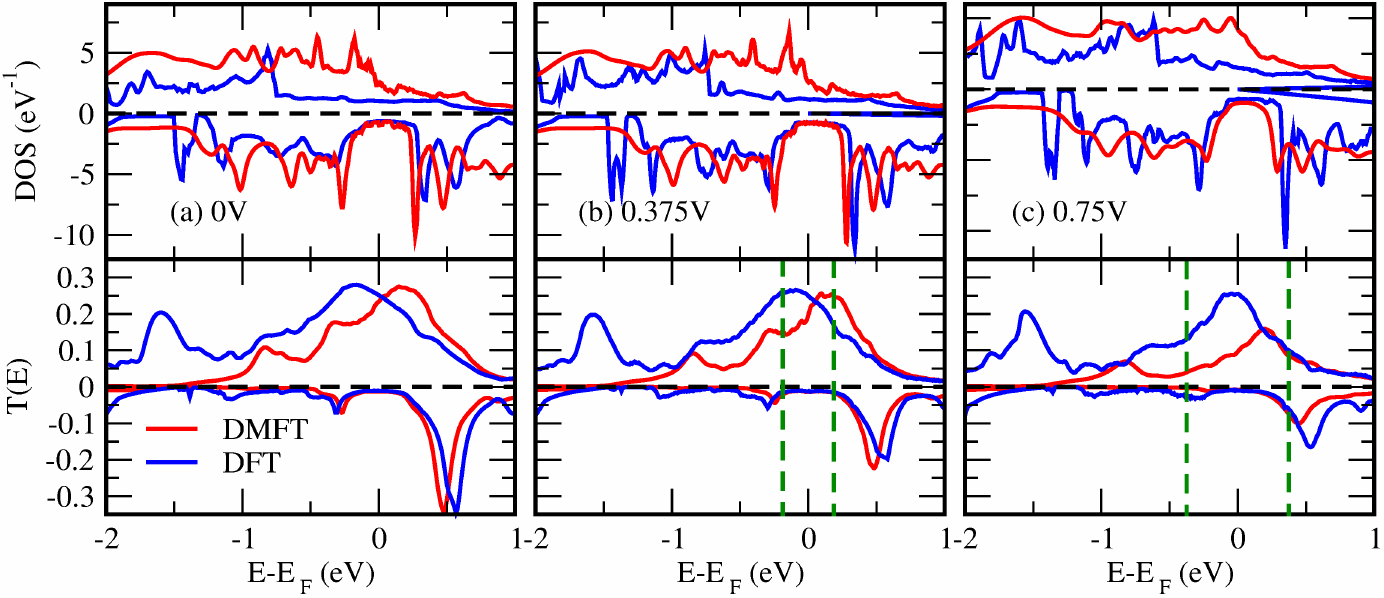}
\caption{Fe $3d$-PDOS and transmission coefficient at (a) 0 V, (b) 0.375 V and (c) 0.75 V, calculated with DFT (red line) 
and DMFT (blue line). Spin-up (down) values are shown positive (negative). The green vertical bars mark the bias window.}
\label{fig:trc} 
\end{figure*}

\begin{figure}[ht] 
\centering
\includegraphics[width=0.45\textwidth]{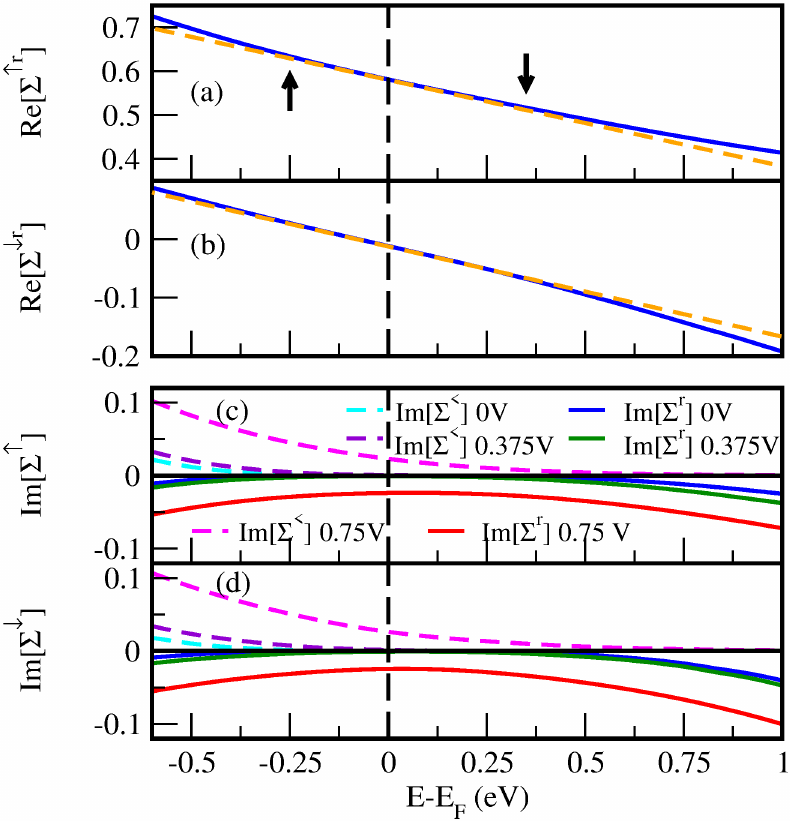} \caption{Self energies as a function of 
energy and bias. Top panels: real part of the retarded DMFT self-energy at zero bias and averaged over the Fe 
$3d$ orbitals, for the (a) spin-up and (b) spin-down channels. The orange line shows a linear fit around $E_\mathrm{F}$, 
with deviations indicating kinks. Bottom panels: the imaginary part of the retarded and lesser self-energy, averaged 
over the Fe $3d$ orbitals, for (c) spin-up and (d) spin-down channels at 0.375 V and 0.75 V.} \label{fig:sigma} 
\end{figure}

Electron-electron interaction is included via a multi-orbital Hubbard-like term, defined only within the correlated 
subspace $\mathcal{C}$ spanned by the Fe $3d$ orbitals. The corresponding NEGF, $\bar{G}^{\sigma}(E,\boldsymbol{k})$, 
is obtained through a projection scheme \cite{andrea_ivan_projection} (see EM). Within DMFT, the many-body 
self-energy of $\mathcal{C}$, is approximated to be local, namely it is momentum independent and diagonal over
the atomic space, with each block $\bar{\Sigma}_{i}^{\sigma}(E)$ representing the $i$-th Fe atom. This is determined 
self-consistently by mapping the  correlated subsystem into an impurity problem, through the condition 
\begin{equation}
 \bar{G}^{\sigma }_{\textnormal{imp}, i}(E) = \bar{G}^{\sigma }_{\textnormal{loc}, i}(E)\:.
\end{equation}
Here, $\bar{G}^{\sigma }_{\textnormal{loc}, i}(E)$ is the $i$-th diagonal block of the local NEGF
\begin{equation}
\bar{G}^{\sigma }_{\textnormal{loc}, i}(E)= \frac{1}{\Omega_{\boldsymbol{k}}} \int_\mathrm{BZ} d\boldsymbol{k}\; \bar{G}^{\sigma }_{\mathcal{C}, i}(E, \boldsymbol{k})\:, \label{LocalGreen}
\end{equation}
where $\Omega_{\boldsymbol{k}}$ is the BZ volume and $\bar{G}^{\sigma }_{\textnormal{imp}, i}(E)$ the impurity 
NEGF. The latter satisfies the Dyson-like equation \cite{ ar.kn.13}
\begin{equation}\label{eq: dyson like equation}
   \bar{G}^{\sigma}_{\textnormal{imp}, i}(E) = \mathcal{G}_{i}^{\sigma}(E) + \mathcal{G}_{i}^{\sigma}(E) \bar{\Sigma}_{\textnormal{imp}, i}^{\sigma}(E)\bar{G}^{\sigma}_{\textnormal{imp}, i}(E)\:,
\end{equation}
where $\mathcal{G}_{i}^{\sigma}(E)$ denotes the dynamical mean-field Green's function (see EM). 
Here, we use a second-order perturbation-theory impurity solver \cite{andrea_sigma_2, andrea_Cu_co}, which provides sufficiently accurate results for metallic systems \cite{andrea_FeO} and is straightforwardly extended to 
non-equilibrium conditions \cite{bookStefanucci}. The method is implemented in the {\sc Smeagol} package 
\cite{ro.ga.06}, which is interfaced with the {\sc Siesta} DFT code \cite{siesta} (see EM for computational details).

\textit{Zero-bias electronic structure.} 
The Fe-projected DOS (PDOS) from DFT and DMFT are shown in Fig. \ref{fig:trc}(a). At zero bias, 
the DFT PDOS exhibits a broad spin-up peak extending over $E_\mathrm{F}$, whereas the spin-down 
PDOS shows a pseudogap of $\sim$0.65~eV, bounded by peaks at $-0.3$~eV and $+0.35$~eV. This 
upper one actually consists of a two-peak structure at 0.35~eV and 0.55~eV, associated 
to inequivalent Fe atoms (see supplementary information of \cite{ha.ne.24}). The PDOS is broaden by 
the hybridization with the leads through the imaginary part of their self-energies. Thus, DFT describes 
Fe$_4$GeTe$_2$ as almost half-metallic~\cite{ha.ne.24}, a property that combined with the high Curie 
temperature makes Fe$_4$GeTe$_2$, as well as related FGT compounds \cite{halder2025}, an interesting 
platform for vdW spintronic devices.

While DFT describes the system in terms of KS single-particle states, DMFT captures many-body electronic excitation. 
Thus, it leads to a redistribution of the PDOS due to the real part of 
the retarded many-body self-energy (averaged over all 
Fe $3d$ orbitals) $\mathrm{Re}[\bar{\Sigma}^{\uparrow\, r}_{\mathrm{avg}}(E)]$, which is positive and varies nearly linearly with negative slope [see Fig. \ref{fig:sigma}(a)]. 
This induces a shift of the spin-up PDOS from lower to higher energies and a concomitant compression, resulting in a slight sharpening of the peak near $E_\mathrm{F}$, where the effective mass is enhanced by a factor of 1.2 with respect to DFT. 
In contrast, the spin-down PDOS is affected much less, 
as $\mathrm{Re}[\bar{\Sigma}^{\downarrow\, r}_{\mathrm{avg}}(E)]$ is an order of magnitude smaller. This leads 
only to a modest reduction of the pseudo-gap at $E_\mathrm{F}$, with the two spin-down peaks moving  
to $\sim$-0.20~eV and 0.25~eV, and an effective mass enhancement of about 1.1. Thus, even in presence of 
dynamical correlations, the system retains its nearly half-metallic character.

Notably, in both spin channels $\mathrm{Re}[\bar{\Sigma}^{\sigma\, r}_{\mathrm{avg}}(E)]$ exhibits kinks at 
approximately -0.2~eV and 0.3~eV, corresponding to the spin-down peaks. This indicates enhanced interaction 
of electrons in one spin channel with electron-hole pairs forming collective excitations in the opposite one. 
Such features are inherently many-body in nature and they have been reported, for instance, at Fe surfaces 
\cite{sc.sc.04}. In what follows, we will demonstrate that such excitations are further amplified in Fe$_4$GeTe$_2$ 
in non-equilibrium conditions, with a pronounced impact on the material's transport properties.

The imaginary part of the DMFT retarded self-energy [Fig. \ref{fig:sigma}(c) and \ref{fig:sigma}(d)] introduces a 
correlation-driven finite lifetime for the electronic excitations, which adds to the hybridization with the leads, 
to yield further broadening and smoothing of the PDOS. Importantly, 
$\mathrm{Im}[\bar{\Sigma}^{\sigma\, r}_{\mathrm{avg}}(E)] \propto -(E - E_\mathrm{F})^2$, namely it exhibits 
Fermi-liquid behaviour \cite{La.57_FL, Vi.22}, indicating that the excitations near $E_\mathrm{F}$ have a 
quasiparticle character and are coherent. Notably, at energies where $\mathrm{Re}[\bar{\Sigma}^{\sigma\, r}_{\mathrm{avg}}(E)]$ 
shows the kinks, $\mathrm{Im}[\bar{\Sigma}^{\sigma\, r}_{\mathrm{avg}}(E)]$ changes slope, becoming 
steeper. This reflects a faster decay of the electronic excitations \cite{Byczuk2007} owing to the interaction 
with the electron-hole pairs. 

The DMFT lesser self-energy [cyan dashed line in Fig. \ref{fig:sigma}(c) and \ref{fig:sigma}(d)] is found to 
accurately satisfy the fluctuation-dissipation theorem \cite{bookStefanucci}, $\mathrm{Im}[\bar{\Sigma}^{\sigma\, <}_{\mathrm{avg}}(E)]=-f(E)\big[\Sigma^{\sigma\, r}_{\mathrm{avg}}(E)-\Sigma^{\sigma\, r}_{\mathrm{avg}}(E)^{\dagger}\big]$, where 
$f(E)$ is the Fermi function, reflecting the occupation of the electronic states in thermal equilibrium. 

\textit{Finite-bias electronic structure.} 
Upon application of a finite bias, DFT predicts a small intra-atomic charge redistribution, leading to a 
slight sharpening of the spin-up PDOS near $E_\mathrm{F}$ and a modest broadening of the spin-down 
one. This reduces slightly the pseudogap [see Fig.~\ref{fig:trc}(b) and \ref{fig:trc}(c)], but this is 
a minor effect. In contrast, DMFT returns a more significant bias-dependent modification 
of the electronic structure, where we can now distinguish different regimes. At low bias, up to 0.25~V, the 
PDOS remains largely unchanged. The bias windows opens across the spin-up (down) peak (pseudogap) 
near the $E_\mathrm{F}$, while the imaginary part of the DMFT self-energy retains its Fermi-liquid character, meaning that the excitations near $E_\mathrm{F}$ remain coherent and quasiparticle-like.

For $V > 0.25$ V, the bias window begins to overlap with the tail of the spin-down valence 
peak [Fig.~\ref{fig:trc}(b)]. At this point, $\mathrm{Im}[\bar{\Sigma}^{\sigma\, r}_{\mathrm{avg}}(E)]$ near $E_\mathrm{F}$ increases, 
indicating faster electronic-excitation damping. Finally, for $V > 0.5$ V, the bias window extends 
over the peak in the spin-down PDOS [Fig.~\ref{fig:trc}(c)] enhancing the coupling between electrons and the collective excitations, marking the onset of a genuinely non-equilibrium, many-body regime. 
The PDOS is significantly reshaped, 
particularly in the spin-up channel, where the features are broadened and smoothened, while the spin-down 
pseudogap gets slightly reduced. This behavior reflects $\mathrm{Im}[\bar{\Sigma}^{\sigma\, r}_{\mathrm{avg}}(E)]$ 
acquiring a relatively large finite value at $E_\mathrm{F}$ [see red line in Fig.~\ref{fig:sigma}(c) and \ref{fig:sigma}(d)], causing electronic excitations to lose their quasiparticle character and becoming increasingly incoherent. Simultaneously, $\bar{\Sigma}^{\sigma\, <}_{\mathrm{avg}}(E)$ [magenta dashed line in Fig.~\ref{fig:sigma}(c) and 
\ref{fig:sigma}(d)] no longer fulfills the fluctuation-dissipation theorem, indicating that electrons no longer follow 
a Fermi-Dirac distribution. 

\textit{Spectral function.} 
The different regimes can be further distinguished by analyzing $A(E, \boldsymbol{k})$. At 
zero bias, DFT gives characteristic parabolic bands crossing $E_\mathrm{F}$ 
near $K$ \cite{ha.ne.24}, with broadening due to the hybridization with the leads [Fig. \ref{fig: spectral_function}(b)]. Nearly flat bands appear 
around $\Gamma$. These have pure $d_{z^2}$ character and are associated with correlated electron and local 
magnetic behavior \cite{Roemer2024}. 

In contrast, the DMFT $A(E, \boldsymbol{k})$ exhibits an upward shift and additional broadening, consistent 
with the PDOS. While at zero-bias, the spectral features around $E_\mathrm{F}$ remain relatively well defined, 
preserving a band-like character [Fig. \ref{fig: spectral_function}(c)], as bias increases and the system enters 
the intermediate non-equilibrium correlated regime, broadening becomes more pronounced across the spectrum 
due to reduced excitation lifetimes. Finally, in the non-Fermi-liquid regime, for instance at $V=0.75$ V 
[see Fig.~\ref{fig: spectral_function}(d)], the bands near $\Gamma$ further flatten and smear into faint traces 
superimposed to an incoherent background.
Similar spectral changes have been reported in high-temperature DMFT calculations~\cite{Xu.Wa.24}. 
By analogy, we interpret our results as arising from voltage-induced energy injection into the Fe$_4$GeTe$_2$ 
monolayer, effectively heating up the electronic system and enhancing scattering processes that destroy 
coherence. Notably, the non-equilibrium $A(E, \boldsymbol{k})$ in current carrying devices is experimentally 
accessible via nano-ARPES techniques \cite{Cu.Jo.23, Ho21, Cu.Sa23}, which offer a promising route to probe 
this ``hot-correlated electron'' regime.

\textit{Bias-dependent transport properties.} 
The signatures of the ``hot-correlated electron'' regime can also be observed in the transport properties. 
The charge current is calculated using the Meir-Wingreen formula \cite{me.wi.92} (see EM). For DFT, this reduces to the finite-bias generalization of the Landauer-B\"uttiker approach \cite{fe.ca.05_2}. 
Transport is coherent and spin conserving, so that the charge current can be decomposed into 
spin-resolved contributions, $I^\sigma$, which are determined by the integral of transmission coefficient, 
$T^\sigma(E)$, within the bias window (see EM). 

$T^\sigma(E)$ follows the Fe-PDOS 
[bottom panels of Fig. \ref{fig:trc}], in particular, for the spin-up channel. In contrast, the spin-down channel 
is dominated by the conduction peak at 0.55~eV, owing to its hybridization with Te states 
and the leads \cite{ha.ne.24}. As the bias window spans the spin-up peak while remaining in the spin-down 
pseudogap, $I^\uparrow$ increases nearly linearly with $V$ [blue line in Fig. \ref{fig: spectral_function}(e)], 
whereas $I^\downarrow$ [blue line in Fig. \ref{fig: spectral_function}(f)] remains an order of magnitude smaller, 
confirming nearly perfect spin filtering up to large bias.

Within DMFT, the current splits into a coherent and an incoherent contribution, 
$I^\sigma = I^\sigma_{\mathrm{C}} + I^\sigma_{\mathrm{IC}}$ (see EM). The coherent part is 
still obtained by integrating $T^\sigma(E)$ over the bias window, although $T^\sigma(E)$ is now 
computed using DMFT. At zero bias, the results reflect the correlation-induced reshaping of the 
Fe-PDOS. As bias increases, however, $T^\sigma(E)$ is suppressed due to the enhanced damping 
of the electronic excitations, yielding a coherent current lower than that computed by DFT; 
$I^\uparrow_{\mathrm{C}}$ even decreases for $V \gtrsim 0.5$ V [Fig.~\ref{fig: spectral_function}(e)]. 

The incoherent contribution, $I^\sigma_{\mathrm{IC}}$ [black line in Fig.~\ref{fig: spectral_function}(e) 
and \ref{fig: spectral_function}(f)] is negligible at low bias, when the monolayer is in the Fermi-liquid regime. 
However, once the Fermi-liquid picture breaks down, $I^\sigma_{\mathrm{IC}}$ increases sharply, reflecting the emergence of incoherent features in the spectral function. This originates from the energy-dependent 
imbalance between electron inflow and outflow due to inelastic scattering processes. In relative magnitude such mechanisms are more important in the spin-down channel, where its onset coincides with an abrupt increase
of the spin-down differential conductance, $d I^\downarrow/dV$ [Fig.~\ref{fig: spectral_function}(g)]. 
This feature can be probed experimentally, through inelastic tunnel spectroscopy, and will provide a 
complementary signature of the hot-correlated electron regime alongside nano-ARPES. Moreover, the 
incoherent contribution reduces the current spin-polarization, and this may have an impact on 2D 
magnetic tunnel junctions. Similar results are also expected for the other FGT compounds, given their 
similar electronic structures \cite{halder2025}.

\textit{Inelastic scattering.} 
The dominant inelastic scattering processes driving $I^\sigma_{\mathrm{IC}}$ 
are similar to those described when modeling inelastic tunneling 
spectroscopy of adatoms~\cite{sc.do.14} and are schematically illustrated in Fig.~\ref{fig: spectral_function}(h) and 
\ref{fig: spectral_function}(i). Specifically, a spin-up electron incoming from the left lead can scatter off an electron in the spin-down band, thereby generating an electron–hole pair excitation that reduces the carrier’s energy.
The process is activated when the bias window becomes as large as the spin-down pseudogap, allowing an electron-hole pair to be generated and enabling the spin-up current-carrying electron to enter an empty state in the right-hand lead (h).  Similarly, a spin-down electron in the conduction band 
can interact with spin-up electron-hole pairs, contributing to the spin-down 
current (i). Beyond their fundamental relevance, these previously overlooked processes in ferromagnets 
may play an important role in current-induced magnetization dynamics, potentially accessible in spin-transfer torque calculations \cite{book1, Maria_STT} replacing DFT with DMFT. 

\textit{Conclusion.} The combination of DFT, DMFT and the NEGF formalism shows that a Fe$_4$GeTe$_2$ monolayer under bias can enter a true non-equilibrium regime, where inelastic 
electron scattering enhances correlations and generates incoherent features in both the excitation spectrum 
and the conductance. These are potentially accessible in experiments. Overall, our results demonstrate that 
our framework captures quantum transport in correlated magnetic materials far beyond the linear response 
limit and the conventional coherent picture, opening a pathway to explore many-body effects in spintronic 
devices.

\section*{Acknowledgments}
The authors thank Liviu Chioncel and Ivan Rungger for useful discussions and for critically reading the 
manuscript. D.N. was supported by the Irish Research Council (Grant No. GOIPG/2021/1468 ). Computational 
resources were provided by Trinity College Dublin Research IT. 

\bibliography{FeMgO}

\newpage

\section*{End Matter}

\textit{Steady-state NEGF formalism.} The NEGF of the central device region is written in block-matrix form as
\begin{equation}\label{eq: GF keldysh representation}
G^{\sigma}(E,\boldsymbol{k}) = \begin{pmatrix}
G^{\sigma\;r}(E,\boldsymbol{k}) & G^{\sigma\;K}(E,\boldsymbol{k}) \\
0 & G^{\sigma\;a}(E,\boldsymbol{k})
\end{pmatrix}.
\end{equation}
where the retarded, advanced, and Keldysh components are
\begin{eqnarray*}
&G^{\sigma\;r}(E,\boldsymbol{k}) = \big[(E+i\eta)S(\boldsymbol{k}) - H_\mathrm{KS}^\sigma(\boldsymbol{k})\\
& - \Sigma^{\sigma\;r}(E, \boldsymbol{k})\big]^{-1}_{\eta \rightarrow 0}\\
&G^{\sigma\;a}(E,\boldsymbol{k}) = [G^{\sigma\;r}(E,\boldsymbol{k})]^\dagger,\\
&G^{\sigma\;K}(E,\boldsymbol{k}) = 2 G^{\sigma\;< (>)} \underset{(-)}{+} [ G^{\sigma\;r}(E,\boldsymbol{k}) - G^{\sigma\;a}(E,\boldsymbol{k})],
\end{eqnarray*}
while the lesser (greater) component is given by
\begin{equation}\label{eq: Glesser}
G^{\sigma\;\lessgtr}(E,\boldsymbol{k}) = G^{\sigma\;r}(E,\boldsymbol{k}) \, \Sigma^{\sigma\;\lessgtr}(E,\boldsymbol{k}) \, G^{\sigma\;a}(E,\boldsymbol{k}).
\end{equation}

These satisfy the relation
\begin{equation}\label{eq: Gr-Ga}
G^{\sigma \; r}(E,\boldsymbol{k})-G^{\sigma \; r}(E,\boldsymbol{k})^{\dagger}= G^{\sigma \; >}(E,\boldsymbol{k})-G^{\sigma \; <}(E,\boldsymbol{k}).
\end{equation}

In the equations above $H^\sigma_{\mathrm{KS}} (\boldsymbol{k})$ is the KS Hamiltonian, $S(\boldsymbol{k})$ 
is the orbital overlap matrix, as the basis orbitals are non-orthogonal, and $\eta$ is an infinitesimal positive number. 
$\Sigma^{\sigma \; r(\lessgtr)}(E, \boldsymbol{k})$ is the retarded (lesser/greater component) of the self-energy, 
which is the sum of the corresponding leads and many-body self-energies, $\Sigma^{\sigma \; r(\lessgtr)}(E, \boldsymbol{k})=\Sigma^{\sigma\; r(\lessgtr)}_\mathrm{L}(E,\mathbf{k})+\Sigma^{\sigma\; r(\lessgtr)}_\mathrm{R}(E,\mathbf{k}) +\Sigma^{\sigma \; r(\lessgtr)}_{\mathrm{MB}}(E)$. $H^\sigma_{\mathrm{KS}}$, $S(\boldsymbol{k})$, the 
NEGFs and the self-energy components are all $N\times N $ matrices, where $N$ is the number of basis orbitals 
in the device central region. In the absence of the many-body self-energy, the above equations reduce to the standard 
KS DFT+NEGF formalism \cite{Ta.Gu.01,Ba.Mo.02,ro.ga.06}
\begin{equation}\label{eq:KS}\begin{split}
&G_\mathrm{KS}^{\sigma\;r}(E,\boldsymbol{k}) =\\ &\big[g^{\sigma\;r}(E,\boldsymbol{k})^{-1} - \Sigma_\mathrm{L}^{\sigma\;r}(E, \boldsymbol{k}) -\Sigma_\mathrm{R}^{\sigma\;r}(E, \boldsymbol{k})\big]^{-1},
\end{split}
\end{equation}
\begin{equation}
\begin{split}
&G^{\sigma\;<}_\mathrm{KS}(E,\boldsymbol{k}) = \\
&G^{\sigma\;r}_\mathrm{KS}(E,\boldsymbol{k}) [\Sigma_\mathrm{L}^{\sigma\;r}(E, \boldsymbol{k}) +\Sigma_\mathrm{R}^{\sigma\;r}(E, \boldsymbol{k})]G^{\sigma\;a}_\mathrm{KS}(E,\boldsymbol{k}).
\end{split}
\end{equation}
with $g^{\sigma\, r} (E,\boldsymbol{k})=[(E+i\eta)S(\boldsymbol{k}) - H_\mathrm{KS}^\sigma(\boldsymbol{k})]^{-1}_{\eta \rightarrow 0}$, the retarded component of $g^{\sigma}(E,\boldsymbol{k})$ introduced in Eq.~(\ref{eq: dyson}).

Since the leads are assumed to remain in local equilibrium, their lesser and greater self-energies can be expressed 
using the fluctuation-dissipation theorem \cite{bookStefanucci},
\begin{equation}\label{eq: leads fd}
\begin{split}
    \Sigma_{\mathrm{L(R)}}^{\sigma\; <}(E,\boldsymbol{k})= & if_{\mathrm{L(R)}}\Gamma^{\sigma}_{\mathrm{L(R)}}(E,\boldsymbol{k})\\
    \Sigma_{\mathrm{L(R)}}^{\sigma\; >}(E,\boldsymbol{k})= & i\big[f_{\mathrm{L(R)}}(E)-1\big]\Gamma^{\sigma}_{\mathrm{L(R)}}(E,\boldsymbol{k}),
\end{split}
\end{equation}  
where 
 $\Gamma^{\sigma}_{\mathrm{L(R)}}(E,\boldsymbol{k})=i\big[\Sigma_{\mathrm{L(R)}}^{\sigma\; r}(E,\boldsymbol{k})-\Sigma_{\mathrm{L(R)}}^{\sigma\; r}(E,\boldsymbol{k})^{\dagger}\big]$ is the coupling of the L (R) lead to the central 
region, and $f_{\mathrm{L(R)}}(E)$ is the Fermi function of the corresponding lead, with chemical potential 
$\mu_\mathrm{L(R)}$. Here, $\Sigma_{\mathrm{L(R)}}^{\sigma\; \lessgtr}$ encode information about the out-of-equilibrium 
in- and outflow of electrons in the central region.

Analogously, for the many-body self-energy we can introduce \cite{andrea_ivan_projection, ne.da.10, book1}
\begin{equation}\label{eq: leads fd MB}
\begin{split}
&\Sigma^{\sigma \; <}_{\textnormal{MB}}(E,\boldsymbol{k})= i F_{\textnormal{MB}}^{\sigma}(E)\Gamma_{\textnormal{MB}}^{\sigma}(E,\boldsymbol{k})\:,\\
&\Sigma^{>}_{\textnormal{MB}}(E,\boldsymbol{k})= i[F_{\textnormal{MB}}^{\sigma}(E)-1]\Gamma_{\textnormal{MB}}(E,\boldsymbol{k})\:.
\end{split}
\end{equation} 
Here, $F_{\textnormal{MB}}^{\sigma}(E)$ denotes the occupation matrix of the central device region. 
In general, it differs from the Fermi function; only in equilibrium does $F_{\textnormal{MB}}^{\sigma}(E)$ 
reduce to the Fermi function. 

\textit{Projection into the correlated subspace.} 
The KS Hamiltonian is projected onto the $\mathcal{C}$ subspace via $\bar{H}_\mathcal{C}^\sigma(\boldsymbol{k})=\;W(\boldsymbol{k})^\dagger H^\sigma_{\mathrm{KS}} (\boldsymbol{k}) W(\boldsymbol{k})$
where $W(\boldsymbol{k})$ is the same transformation matrix defined in Eqs. (17) and (A17) of 
Ref. \cite{andrea_ivan_projection}. This transformation also orthogonalizes the orbitals within $\mathcal{C}$. 
The projected Hamiltonian $\bar{H}_\mathcal{C}^\sigma(\boldsymbol{k})$ is then supplemented by a 
Hubbard-like interaction term. The NEGF of the $\mathcal{C}$ subspace, 
$\bar{G}_{\mathcal{C}}^{\sigma}(E,\boldsymbol{k})$, can be defined in a form analogous to 
Eq.~(\ref{eq: GF keldysh representation}) with components
\begin{equation}
    \bar{G}_{\mathcal{C}}^{\sigma}(E,\boldsymbol{k})   = \begin{pmatrix}
        \bar{G}_\mathcal{C}^{\sigma\;r}(E,\boldsymbol{k}) &   \bar{G}_\mathcal{C}^{\sigma\;K}(E,\boldsymbol{k}) \\
        0 &   \bar{G}_\mathcal{C}^{\sigma\;r}(E,\boldsymbol{k})^\dagger
            \end{pmatrix}.
\end{equation}
The projected self-energy is given by $\bar{\Sigma}^{\sigma \; r(<)}(E, \boldsymbol{k})=\bar{\Sigma}^{\sigma\; r(<)}_\mathrm{L}(E,\mathbf{k})+\bar{\Sigma}^{\sigma\; r(<)}_\mathrm{R}(E,\mathbf{k}) +\bar{\Sigma}^{\sigma \; r(<)}_{\mathcal{C}}(E)$. 
The new L (R) lead self-energy for $\mathcal{C}$, $\bar{\Sigma}^{\sigma\; r(<)}_\mathrm{L(R)}(E,\mathbf{k})$ is 
obtained as detailed in Ref. \cite{andrea_ivan_projection} [see Eqs. (71) and (72)]. The many-body self-energy 
matrix of $\mathcal{C}$ in DMFT is both $\boldsymbol{k}$-independent and diagonal in the atomic sites,
namely
\begin{equation}\label{eq: block structure of mb self energy}
 \bar{\Sigma}^{\sigma \;  r(<)}_{\mathcal{C}}(E)= 
 \left( \begin{array}{cccc}
 \bar{\Sigma}^{\sigma\;  r(<)}_{1}(E) &   0 & ... & 0          \\
 0 & \bar{\Sigma}^{\sigma\;  r(<)}_{2}(E) & ... & 0           \\
 0 & 0 & ... & \bar{\Sigma}^{\sigma\;  r(<)}_{N_{\textnormal{Fe}}}(E)           \\
\end{array} \right),
\end{equation}
where $N_{\textnormal{Fe}}$ is the number of Fe atoms in the central region (equal to four in our case), 
and each block $\bar{\Sigma}^{\sigma \; r(<)}_i(E)$ represents the retarded (lesser) DMFT self-energy 
for atomic site $i$ within the correlated subspace $\mathcal{C}$.
Finally, the DMFT self-energy within $\mathcal{C}$ can be projected back to the central region 
as \cite{andrea_ivan_projection}  
\begin{equation}
    \Sigma^{\sigma \; r(<)}_{\textnormal{MB}}(E,\boldsymbol{k})=W(\boldsymbol{k})^{-1}\bar{\Sigma}^{\sigma \; r(<)}_{\mathcal{C}}(E)[W(\boldsymbol{k})^{\dagger}]^{-1}.
\end{equation}

\textit{Dynamical mean-field.} 
The retarded and advanced components of the dynamical mean-field in Eq.~(\ref{eq: dyson like equation}) 
are given by
\begin{eqnarray} \label{eq: dynamical retarded mean field} 
&\mathcal{G}^{\sigma \; r}_i(E) = \Big[G^{\sigma \; r}_{\textnormal{loc}, i}(E)^{-1} + \bar{\Sigma}_{i}^{\sigma \; r}(E)\Big]^{-1},
 \\
 &\mathcal{G}^{\sigma \; <}_{i}(E) = \mathcal{G}^{\sigma \; r}_{i}(E) \Delta_{i}^{<}(E) \mathcal{G}^{\sigma \; a}_{ i}(E), 
\end{eqnarray} 
where $\mathcal{G}^{\sigma \; a}_{i}(E) = [\mathcal{G}^{\sigma \; r}_{i}(E)]^\dagger$
and $\Delta_{i}^{<}(E)$ is the lesser hybridization function, defined as
\begin{equation} \label{eq. lesser hybridisation}
\Delta_{i}^{<}(E) = \bar{G}^{\sigma \; r}_{\textnormal{loc}, i}(E)^{-1} \bar{G}^{\sigma \; <}_{\textnormal{loc}, i}(E) \bar{G}^{\sigma \; a}_{\textnormal{loc}, i}(E)^{-1}  - \bar{\Sigma}_{i}^{\sigma \; <}(E).
\end{equation}
For the first iteration of the DMFT self-consistent loop, the DMFT self-energies 
$\bar{\Sigma}^{\sigma \; r}_i(E)$ and $\bar{\Sigma}^{\sigma \; <}_i(E)$ are initialized to zero.
At zero bias, namely in equilibrium, $\mathcal{G}^{\sigma \; <}_i(E)$ is related to $\mathcal{G}^{\sigma \; r}_i(E)$ 
via the fluctuation-dissipation theorem, and the DMFT equations reduce to those of layered DMFT in 
Refs.~\cite{valli_dmft, andrea_Cu_co}.

\textit{Charge current.} 
The two-terminal steady-state charge current is given by the Meir-Wingreen formula \cite{me.wi.92}
\begin{equation}\label{eq: symmetrised MW}
    \begin{split}
            I^{\sigma} = &\frac{e}{2h\Omega_{\boldsymbol{k}}} \int  d\boldsymbol{k}  \int dE  \;\textnormal{Tr}\Big[[\Sigma_{\textnormal{L}}^{\sigma \;<}(E,\boldsymbol{k})-\Sigma_{\textnormal{R}}^{\sigma \;<}(E,\boldsymbol{k})]\times \\
            &\;\;\;\;\;\;\;\;\;\;\;\;\;\;\;\;\;\;\;\;\big[G^{\sigma \;r}(E,\boldsymbol{k})
             -G^{\sigma \; a}(E,\boldsymbol{k})\big]\\ & \;\;\;\;\;\;\;\;\;\;\;\;\;\;\;\;+i[\Gamma_{\textnormal{L}}^{\sigma}(E,\boldsymbol{k})-\Gamma^{\sigma}_{\textnormal{R}}(E,\boldsymbol{k})]G^{\sigma \;<}(E,\boldsymbol{k})\Big].
    \end{split}
\end{equation}
Using Eq. (\ref{eq: Gr-Ga}) together with the expressions for the lesser and greater components of the lead 
and many-body self-energies in Eqs.~(\ref{eq: leads fd}) and (\ref{eq: leads fd MB}), the current can be 
rewritten as the sum of the coherent and incoherent contributions:
\begin{equation} \label{eq: coherent + incoherent}
    I^{\sigma} = I_{\textnormal{C}}^{\sigma}+ I_{\textnormal{IC}}^{\sigma}.
\end{equation}

The coherent contribution reads
 \begin{equation} 
\begin{split}
    I_{\textnormal{C}}^{\sigma} = \frac{e}{h}  \int_{-\infty}^{\infty}\! dE \;
 T^\sigma(E) \big[f_{\textnormal{L}}(E)-f_{\textnormal{R}}(E)\big],\label{eq: coherent current}
\end{split}
\end{equation}
where $T^{\sigma}(E)$ denotes the  transmission coefficient, given by
\begin{equation}\begin{split}
T^\sigma(E)= \frac{1}{\Omega_{\boldsymbol{k}}}\int d\boldsymbol{k}\;\textnormal{Tr} & \Big[\Gamma^{\sigma}_{L}(E, \boldsymbol{k})  G^{\sigma\; r}(E, \boldsymbol{k})\\
 &\Gamma^{\sigma}_{R}(E, \boldsymbol{k})G^{\sigma\; r}(E, \boldsymbol{k})^{\dagger}\Big].
\end{split}\label{eq: TRC many body k-dependent}
\end{equation} 
Equation (\ref{eq: coherent current}) is the generalization of the Laudauer-B\"uttiker 
formula to finite bias (see, for example, reference \cite{Da.95,sanvito, ro.ga.06}).

The incoherent contribution is
\begin{equation} \label{eq: incoherent current k dependent}
    \begin{split}
        I_{\textnormal{IC}} = &\frac{e}{2h\Omega_{\boldsymbol{k}}}\int d\boldsymbol{k}\int dE\; \textnormal{Tr}\Big\{\big[f_{\textnormal{L}}(E)-f_{\textnormal{MB}}^{\sigma}(E)\big]\Gamma_{\textnormal{MB}}^{\sigma}(E)\\ &\times G^{\sigma\; r}(E, \boldsymbol{k})^{\dagger}\Gamma^{\sigma}_{L}(E, \boldsymbol{k}) G^{\sigma\; R}(E, \boldsymbol{k}) \Big\} \\
        & - \textnormal{Tr}\Big\{\big[f_{\textnormal{R}}(E)-f_{\textnormal{MB}}^{\sigma}(E)\big]\Gamma_{\textnormal{MB}}^{\sigma}(E)G^{\sigma\; r}(E, \boldsymbol{k})^{\dagger}\\
        & \times \Gamma^{\sigma}_{R}(E, \boldsymbol{k}) G^{\sigma\; r}(E, \boldsymbol{k})\Big\}.
    \end{split}
\end{equation}
In the KS single-particle description, the many-body self-energy is absent, and thus the incoherent contribution 
vanishes. In this case, the current is given solely by Eq.~(\ref{eq: coherent current}), with $T^{\sigma}(E)$ evaluated 
using the KS NEGF, Eq. (\ref{eq:KS}). 

In DMFT, both contributions can be non-negligible. $T^{\sigma}(E)$ incorporates the NEGF with 
the DMFT self-energy, and represents the transmission for the electronic excitations renormalized in 
both energy and lifetime by dynamical correlation effects. The incoherent contribution instead captures 
the impact of inelastic scattering on transport. Notably, Eq. (\ref{eq: incoherent current k dependent}) 
resembles Eq. (\ref{eq: coherent current}), but with $f_{\textnormal{MB}}^{\sigma}$ and $\Gamma^{\sigma}_{\textnormal{MB}}$
replacing $f_\mathrm{R}$ ($f_\mathrm{L}$) and $\Gamma^{\sigma}_{\textnormal{R}}$ ($\Gamma^{\sigma}_{\textnormal{L}})$ 
in the first (second) term of the integrand. This analogy suggests an intuitive picture in which inelastic scattering 
is effectively represented by an additional ``virtual electrode''. 
$I_{\textnormal{IC}}$ can then be interpreted as the current flowing from the left lead into this virtual 
electrode and subsequently emerging into the right lead with modified energy and phase. This picture 
closely resembles the phenomenological concept of the dephasing probe introduced by B\"uttiker \cite{bu.86}, 
commonly used to describe the crossover between coherent and non-coherent transport in nano-conductors. 
Eqs. (\ref{eq: incoherent current k dependent}) and (\ref{eq: coherent current}) with the DMFT self-energy 
provide a quantitative, atomistic, and theoretically grounded framework to address this crossover.

\textit{Computational details.} 
DFT+NEGF calculations are performed using the same computational details as in Ref. \cite{ha.ne.24}.
DMFT calculations are carried out with orbital-dependent screened Coulomb interaction parameters for 
the Fe $3d$ orbitals, expressed in terms of the Slater integrals $F^0$, $F^2$, and $F^4$ \cite{im.fu.98}. 
The ratio $F^4/F^2$ is fixed to the atomic value $\approx 0.625$ \cite{an.gu.91}. The average interaction 
parameters are defined as $U = F^0$ and $J = (F^2+F^4)/14$, and in this work, take the values 
$U = 1.5$ eV and $J = 0.5$ eV.

The treatment of double-counting corrections remains an open issue for transport calculations. For a 
practical implementation, we assume that the double-counting term exactly cancels the first-order 
self-energy contribution from the perturbative impurity solver (both corresponding to ``on-site'' energy shifts), 
and neglect both. At zero bias, this approximation yields results similar to those in Ref. \cite{ha.ne.24}, where 
the first-order term and a double-counting correction in the localized limit \cite{andrea_sigma_2} were included, 
provided that we use the reduced interaction parameter $U = 1.5$ eV mentioned above in place of the $U = 3$~eV 
of Ref. \cite{ha.ne.24}.

The local Green's function is computed using a $20 \times 20$ $\boldsymbol{k}$-mesh. DMFT calculations 
are performed directly on the real energy axis with a grid of 10,000 points spanning the range $[-25,15]$ eV. 
The resulting converged DMFT self-energy is then used to evaluate the DOS and transmission coefficient 
on a finer $80 \times 80$ $\boldsymbol{k}$-mesh.

\end{document}